\documentclass[12pt,onecolumn,draftclsnofoot]{IEEEtran}

\usepackage{graphicx}
\usepackage{graphics}
\usepackage{epsfig}

\newcommand{\beq}{\begin{equation}}
\newcommand{\enq}{\end{equation}}
\newcommand{\beqa}{\begin{eqnarray}}
\newcommand{\enqa}{\end{eqnarray}}
\newcommand{\beqan}{\begin{eqnarray*}}
\newcommand{\enqan}{\end{eqnarray*}}

\newcommand{\bit}{\begin{itemize}}
\newcommand{\eit}{\end{itemize}}

\newcommand{\mycal}[1]{\mathcal #1}
\newcommand{\comment}[1]{}

\begin{document}

\title{\Large\bfseries Is the cyclic prefix necessary?}
\author{Naresh~Sharma and Ashok~Armen~Tikku
\thanks{Corresponding author: Ashok Armen Tikku.
The material in this paper was presented in part at the
International Symposium on Information Theory (ISIT), Seattle, WA,
USA, July 2006.
N. Sharma is with the Indian Institute of Technology (IIT),
New Delhi 110016, India,
and A.A. Tikku is with the Columbia University, New York, NY 10027, USA.
(email: nareshs@ieee.org, aat2119@columbia.edu).
}
}

\maketitle

\begin{abstract}

We show that one can do away with the cyclic prefix (CP)
for SC-FDE and OFDM at the cost of a moderate increase in  the
complexity of a DFT-based receiver. Such an approach effectively
deals with the decrease in the number of channel uses due to the
introduction of the CP. It is shown that the SINR for SC-FDE remains
the same asymptotically with the proposed receiver without CP as that of the
conventional receiver with CP. The results are shown for $N_t$ transmit
antennas and $N_r$ receive antennas where $N_r \geq N_t$.

\end{abstract}

\section{Introduction}

Inter-symbol Interference (ISI) introduced by the time-varying multi-path channel is one
of the main limiting factors for high speed data communications.
Discrete Fourier Transform (DFT) based receiver used in SC-FDE (single-carrier
frequency-domain equalization) and OFDM (orthogonal frequency-division
multiplexing) offer a low complexity alternative to deal with this problem
(see for example \cite{pro,falconer,digg} and references therein).

Fundamentally, these techniques rely on the redundancy in the
form of CP that with appropriately chosen length makes the linear convolution
introduced by the physical channel look like a circular convolution that can be
dealt well with DFT or Inverse DFT (IDFT). This redundancy however results
in decrease of the number of channel uses available for signal transmission.


We will show that without CP and using a receiver that is designed to work
with cyclic prefix, one suffers from an interference component
at the start and the end of the DFT frame. This edge effect decays
exponentially as one moves inside
the DFT frame unless the poles of a certain inverse filter are on the
unit circle, in which case, the interference is present for the entire frame.

We show that it is guaranteed that for the MMSE SC-FDE, the poles
will never lie on the unit circle and hence the edge effect will truly be at
the edges assuming that the DFT frame is large enough so that the symbols at the
interior of the frame are unaffected due to the exponential decay. Hence
by discarding the symbols at the edges and re-processing the discarded
symbols by putting them in the interior of another frame, one can
obtain a performance that can made as close as possible to
the one obtained with CP.

We define the notation as used throughout this paper:
\bit
\item $\{ y_m \}$ denotes a finite length sequence of vectors indexed by $m$ whose length is specified
in the context, where $y_m$ is a $N_t \times 1$ vector whose $k$th element is denoted by $y_{m,k}$.
\item DFT of a $N$-point vector-sequence is denoted by $\{\tilde{y}_n\} = $ DFT$\{y_m\}$ 
and is given by
\beq
\label{eqdft}
\tilde{y}_n = \frac{1}{\sqrt{N}} \sum_{m=0}^{N-1} y_m e^{-j2 \pi mn/N}.
\enq
Note that DFT is element-wise  and one can write (\ref{eqdft}) as
\beq
\tilde{y}_{n,k} = \frac{1}{\sqrt{N}} \sum_{m=0}^{N-1} y_{m,k} e^{-j2 \pi mn/N}.
\enq
\item Inverse DFT (IDFT) of a $N$-point
vector-sequence is denoted by $\{y_m\} = $ IDFT$\{\tilde{y}_n\}$ and is given by
\beq
y_m = \frac{1}{\sqrt{N}} \sum_{n=0}^{N-1} \tilde{y}_n e^{j2 \pi mn/N}.
\enq
\item ${\cal CN}(0,\sigma^2)$ denotes a circularly symmetric complex Gaussian random
variable with zero mean and variance of $\sigma^2$.
\item Let $x_m = $ $\{h_m \otimes u_m \}$
$\stackrel{\Delta}{=}
\sum_{q=0}^{N-1} h_q u_{(m-q) \mbox{\scriptsize mod} N}$
denote the circular convolution
of the two $N$-point sequences $\{h_m\}$ and $\{u_m\}$, where
$h_m$ is a $N_r \times N_t$ matrix and $u_m$ is a $N_t \times 1$ vector. Note
that if $\{\tilde{x}_n\} = \mbox{DFT}\{x_m\}$, $\{\tilde{h}_n\} = \mbox{DFT}\{h_m\}$,
and $\{\tilde{u}_n\} = \mbox{DFT}\{u_m\}$, then
$\tilde{x}_n = \sqrt{N} \tilde{h}_n \tilde{u}_n$.
\item $u^*$ denotes the conjugate of $u$.
\item $||A||_\infty$ denotes the $L$-infinity norm of matrix $A$ given by
$||A||_\infty = \max_{i,j} |A_{i,j}|$.
\item E$\{\cdot\}$ denotes the expectation.
\item $x^\dagger$ denotes the conjugate transpose of $x$.	
\eit

\section{System Model}

Let $\{u_m\}$, $m=-\infty,\cdots,\infty$, $u_m \in {\mathcal C}^{N_t}$
be the input to a $L$-tap channel given by
\[ {\bf \underline{h}} = [h_0,\cdots,h_{L-1}], \]
where $h_i$'s are $N_r \times N_t$ matrices.
If $x_m$ and $\nu_m$ denote the channel output and noise respectively
at time instant $m$, then the system model is given by
\beq
\label{sysmod}
x_m = \sum_{l=0}^{L-1} h_l u_{m-l} + \nu_m.
\enq
We will assume that the channel is perfectly known at the receiver, is
quasi-static that remains constant over a frame and changes independently
from one frame to another, whose elements are ${\cal CN}(0,1)$, and the channel
random process is spatially and temporally white,
$\nu_m$ is i.i.d. whose elements are uncorrelated and each element is
${\cal CN}(0,N_0)$, and without loss of generality that
the average transmitted power is unity i.e. E$\{||u_m||^2\}=1$.

In what follows, we consider OFDM, CP SC-FDE, CP-less SC-FDE,
and CP-less OFDM with the CP present in the first two techniques.

We will divide the transmitted symbols into frames of length $N+C$ each,
where $N$ denotes the number of symbols carrying information and $C$ denotes
the number of redundant symbols due to CP. It is well-known that for $C \geq L-1$,
the linear convolution will be the same as the circular convolution after discarding
the first $C$ samples of the received signal. Let the information carrying
symbols be denoted by
\beq
\tilde{{\bf \underline{y}}} = [\tilde{y}_0,\cdots,\tilde{y}_{N-1}].
\enq
For OFDM only, these symbols undergo an additional transformation by using the IDFT as
\beq
\label{ofdmtx}
\{y_m\} = \mbox{IDFT}\{\tilde{y}_n\}.
\enq
For CP SC-FDE and CP-less SC-FDE,
\beq
y_m = \tilde{y}_m, ~~~ \forall ~~~ m.
\enq

The transmitted frame of size $N_t \times (N+C)$ is
transmitted during the time instants from
$-C,-C+1,\cdots,0,1,\cdots,N-1$ and is given by
\beq
{\bf \underline{u}} = [\stackrel{\mbox{\scriptsize Cyclic Prefix}, C}
{\overbrace{{y_{N-C},y_{N-C+1},\cdots,y_{N-1}}}},
\stackrel{\mbox{\scriptsize Data}, N}{\overbrace{y_0,y_1,\cdots,y_{N-1}}}].
\enq

For the CP-less case, $C=0$ and hence the transmitted signal in time
instants $-C,\cdots,N-1$ is given by
\beq
{\bf \underline{u}} = [\stackrel{\mbox{\scriptsize From previous frame}, C}
{\overbrace{{w_{N-C},\cdots,w_{N-1}}}},
\stackrel{\mbox{\scriptsize Data}, N}{\overbrace{y_0,\cdots,y_{N-1}}}],
\enq
where $w_m$'s are the transmitted vectors from the previous frame.

\section{Receiver}

The channel output is recorded at the time instants $m = 0,\cdots,N-1$ and
we take the $N$-point DFT given by
\beq
\{\tilde{r}_n\} = \mbox{DFT}\{x_{m}\}.
\enq

\subsection{OFDM Receiver}

For the OFDM receiver, using (\ref{ofdmtx}) we have
\beqa
\tilde{r}_n & = & \frac{1}{\sqrt{N}} \sum_{q=0}^{N-1} x_{q} e^{-j2\pi qn/N} \nonumber \\
& = & \frac{1}{\sqrt{N}} \sum_{q=0}^{N-1} \left(\sum_{l=0}^{L-1} h_l u_{q-l} + \nu_{q} \right) e^{-j2\pi qn/N} \nonumber \\
& = & \frac{1}{\sqrt{N}} \sum_{q=0}^{N-1} \left(\sum_{l=0}^{L-1} h_l y_{(q-l) \mbox{\scriptsize mod} N}
+ \nu_{q} \right) e^{-j2\pi qn/N} \nonumber \\
\label{ofdmrx}
& = & \sqrt{N} \tilde{h}_n \tilde{y}_n + \tilde{\nu}_n,
\enqa
where $\{ \tilde{\nu}_n\} = \mbox{DFT}\{\nu_{q} \}$ and
$\{ \tilde{h}_n\} = \mbox{DFT} \{h_l\}$.
Since DFT is unitary, $\{\tilde{\nu}_n\}$ has the same statistics as $\{\nu_m\}$.

\subsection{CP SC-FDE Receiver}

Following the analysis for the OFDM receiver, we arrive at (\ref{ofdmrx}). Note that unlike OFDM
the information bearing signals for
this case are $y_m$'s. We multiply (\ref{ofdmrx}) by a $N_t \times N_r$ matrix denoted by
$\tilde{g}_n$ and then take the IDFT to get
\beqa
\hat{y}_m & = & \frac{1}{\sqrt{N}} \sum_{n=0}^{N-1} \tilde{g}_n \tilde{r}_n 
e^{j2\pi mn/N} \nonumber \\
& = & \frac{1}{\sqrt{N}} \sum_{n=0}^{N-1} \tilde{g}_n \tilde{h}_n \tilde{y}_n
e^{j2\pi mn/N} + \nonumber \\
& & ~~~~~~~~
\frac{1}{\sqrt{N}} \sum_{n=0}^{N-1} \tilde{g}_n \tilde{\nu}_n e^{j2\pi mn/N} \nonumber\\
\label{tmp1}
& \stackrel{a}{=} & \frac{1}{\sqrt{N}} \sum_{k=0}^{N-1} p_k y_{(m-k) \mbox{\scriptsize mod} N} +
\nonumber \\
\label{sc-fde_rx}
& & ~~~~~~~~
\frac{1}{\sqrt{N}} \sum_{n=0}^{N-1} \tilde{g}_n \tilde{\nu}_n e^{j2\pi mn/N},
\enqa
where in a, $\tilde{p}_n = \tilde{g}_n \tilde{h}_n$, and
\beq
\{p_m\} = \mbox{IDFT}\{\tilde{g}_n \tilde{h}_n\}.
\enq
Note that for the zero-forcing (ZF) SC-FDE,
\beq
\tilde{g}_n = {1 \over \sqrt{N}}
(\tilde{h}_n^\dagger \tilde{h}_n)^{-1} \tilde{h}_n^\dagger,
\enq
and for the MMSE SC-FDE,
\beq
\label{mmsefde}
\tilde{g}_n =
\sqrt{N}(N \tilde{h}_n^\dagger \tilde{h}_n + N_0 I)^{-1} \tilde{h}_n^\dagger.
\enq
We note that in (\ref{mmsefde}), we are first doing the ZF or MMSE reception
in the frequency domain and then taking the DFT.
Since DFT is a unitary operation, it can be easily shown that ZF or MMSE in
time domain can be written as a concatenation of ZF or MMSE respectively
in frequency domain followed by DFT. This is shown in the appendix.

\comment{The computation of signal to interference plus noise ratio (SINR) may look
formidable for this case, but it can be simplified as follows. The noise
term arising due to channel noise has a variance
of $N_0 \sum_{q=0}^{N-1} \left|G_n/\tilde{h}_n\right|^2/N$. The channel gain
for the signal part is given by $g_0/\sqrt{N}$ and the interference power
from other symbols is given by
\beqa
\frac{1}{N} \left[ \sum_{m=0}^{N-1} |g_m|^2 - |g_0|^2 \right] & = & \sum_{n=0}^{N-1} \frac{G_n^2}{N} -
\left( \sum_{n=0}^{N-1} \frac{G_n}{N} \right)^2 \nonumber \\
\label{tmp2}
& = & \sigma_G^2,
\enqa
where we have used the Parseval's Theorem $\sum_{m=0}^{N-1} |g_m|^2 = \sum_{n=0}^{N-1} G_n^2$,
$g_0 = \sum_{n=0}^{N-1} G_n/\sqrt{N}$, the fact that $G_n$ is real for all $n$, and
$\sigma_G^2$ denotes the variance of the sequence $\{G_n\}$.
The SINR for any symbol is given by
\beq
\mbox{SINR}_{\mbox{\scriptsize MMSE}} =
\frac{\left(\sum_{n=0}^{N-1} G_n\right)^2}{N \left( N\sigma_G^2 + N_0 \sum_{q=0}^{N-1} G_n^2/|\tilde{h}_n|^2 \right) }.
\enq
For the case of ZF SC-FDE, one can simplify the above expression by noting that
$\sigma_G^2=0$ and substituting $G_n = 1$ to get
\beq
\mbox{SINR}_{\mbox{\scriptsize ZF}} = \frac{N}{N_0 \sum_{q=0}^{N-1} 1/|\tilde{h}_n|^2 }.
\enq
}

\subsection{CP-less SC-FDE}

The receiver is kept the same as the CP SC-FDE except that one could vary the
DFT frame size. In this case, the
absence of CP doesn't make the linear convolution as a circular convolution and
there is a spill-over of the signals from the previous frame causing interference.
As we shall see below, for the MMSE case in particular, it is guaranteed that the
interference levels will fall exponentially as one moves inside the frame from either of the
two ends. Because of this phenomenon, one can discard the symbols at either ends of
the frame and declare only the interior symbols of the frame as the equalized symbols.
The discarded symbols at either end can be equalized by putting them on the interior of
a different frame. This indicates that having sliding and over-lapping frames (with
the overlap determined by the rate of decay of the interference levels) will result in
equalization of all symbols. It also follows that the additional interference due
to lack of CP can be \emph{controlled} by increasing the number of discarded symbols.
Hence by increasing the receiver complexity, one can  compensate
for the absence of CP and also increase the effective number of channel uses as compared
to the CP-based transmission.

We take the $N$-point DFT of the received sequence as in OFDM/CP SC-FDE to get
\beqa
\tilde{r}_n & = & \frac{1}{\sqrt{N}} \sum_{q=0}^{N-1} x_{q} e^{-j2\pi qn/N} \nonumber \\
\label{cpless1}
& = & \sqrt{N} \tilde{h}_n \tilde{y}_n + \tilde{\nu}_n + \tilde{\kappa}_n,
\enqa
where
\beq
\{\tilde{\kappa}_n\} = \mbox{DFT}\left\{ \sum_{l=q+1}^{L-1} h_l \left( w_{N+q-l} - y_{N+q-l} \right) \right\}.
\enq
We note that the last summation term in (\ref{cpless1}) is the only difference between CP-less case and the schemes with
CP. We multiply
$\tilde{r}_n$ in (\ref{cpless1}) by $\tilde{g}_n$ and then take the IDFT to get
\beqa
\hat{y}_m & = & \frac{1}{\sqrt{N}} \sum_{n=0}^{N-1} \tilde{g}_n \tilde{r}_n 
e^{j2\pi mn/N} \nonumber \\
& = & \frac{1}{\sqrt{N}} \sum_{k=0}^{N-1} p_k y_{(m-k) \mbox{\scriptsize mod} N} +
\nonumber \\
& & ~~~~~~~~
\frac{1}{\sqrt{N}} \sum_{n=0}^{N-1} \tilde{g}_n \tilde{\nu}_n e^{j2\pi mn/N}
+ \xi_m,
\enqa
where $\tilde{p}_n = \tilde{g}_n \tilde{h}_n$,
\beq
\{p_m\} = \mbox{IDFT}\{\tilde{g}_n \tilde{h}_n\},
\enq
\beq
\label{xieq}
\xi_m = \sum_{q=0}^{L-2} \sum_{l=q+1}^{L-1}
\gamma_{(m-q) \mbox{\scriptsize mod} N} h_l \left( w_{N+q-l} - y_{N+q-l} \right) ,
\enq
and $\{\gamma_m\} = \mbox{IDFT} \{\tilde{g}_n\}/\sqrt{N}$ i.e.
\beq
\label{eqgamma}
\gamma_m = \frac{1}{N} \sum_{n=0}^{N-1} \tilde{g}_n e^{j2\pi mn/N}.
\enq

Let us assume that it is possible to choose a $D > L-1$ and $D \leq m \leq N-D$, such
that $||\gamma_m||_\infty < \epsilon$, for any $\epsilon > 0$.
Then for $D \leq m \leq N-D$, one can upper bound $||\xi_m||$ as
\beqa
|\xi_m| & \leq & \sum_{q=0}^{L-2} \sum_{l=q+1}^{L-1} N_r N_t
\beta_1 \beta_2 \beta_3 \epsilon \\
\label{ubxi}
& \leq & {(L-2)(L-1) N_r N_t \beta_1 \beta_2 \beta_3 \epsilon \over 2},
\enqa
where $\beta_1 = \max_q ||\gamma_{(m-q) \mbox{\scriptsize mod} N}||_\infty$,
$\beta_2 = \max_l ||h_l||_\infty$ and $\beta_3 = ||w_{N+q-l} - y_{N+q-l}||_\infty$.

Hence by choosing $\epsilon$ small enough, one can make the interference term
$\xi_m$ that arises due to the absence of
CP negligible. This makes the SINR of the symbols in the interior of the frame the
same as the corresponding symbols where the CP is present.

We now show that it is indeed possible to have $||\gamma_m||_\infty \rightarrow 0$ in
the interior of a large enough frame i.e. when $D \leq m \leq N-D$ and $N$ is large if
the poles of a certain inverse filter are not on the unit circle. Let us write
\beq
\tilde{g}_n = \left[H(z)H^\dagger(1/z^*) + K\right]^{-1} H^\dagger(1/z^*),
\enq
where $H(z)$ is matrix function of scalar variable $z$ given by
\beq
H(z) = \sum_{l=0}^{L-1} h_l z^{l},
\enq
$z = e^{-j2\pi n/N}$, $K = 0$ for the
ZF SC-FDE, and $K=N_0$ for the MMSE SC-FDE.
Note that for $z \neq 0$, we can write
\beq
\label{eqgn}
\tilde{g}_n = \left[z^{L-1} H(z)H^\dagger(1/z^*) + K z^{L-1}\right]^{-1} z^{L-1} H^\dagger(1/z^*).
\enq
Let us define
\beq
R(z) \stackrel{\Delta}{=} \det\left( z^{L-1}H(z)H^\dagger(1/z^*) + Kz^{L-1} \right).
\enq
We note that $R(z)$ is self-reciprocal since
\beq
R(z) = z^{2L-2} R^*(1/z^*).
\enq
As a consequence if $\phi$ ($|\phi| \neq 0$) is a root of $R(z)$ i.e. $R(\phi) = 0$, then $R(\phi) = \phi^{2L-2} R^*(1/\phi^*) = 0$,
or $1/\phi^*$ is also a root of $R(z)$. Hence we can write
\beqa
R(z) & = & c \prod_{k=1}^{N_r(2L-2)} (z - \beta_k)(z - 1/\beta_k^*),
\enqa
where $c$ is a constant dependent on the channel.
Note that any element of $\tilde{g}_n$ in (\ref{eqgn})
will have $R(z)$ in the denominator,
and furthermore, the degree of the numerator (i.e. the highest
power of $z$) will be smaller than $R(z)$.
Hence the partial fraction expansion of the $(i,j)$ element of
$\tilde{g}_n$ can be written as
\beq
\tilde{g}_n^{i,j} = \sum_{l=1}^P \frac{\alpha_l^{i,j}}{z - \beta_l},
\enq
where $z = e^{-j2\pi n/N}$, $P = 2N_r(L-1)$ for $N_r > 1$ and $P=L-1$ for $N_r=1$
for the ZF SC-FDE, and we have assumed that all the $\beta_l$'s are distinct, an event
that occurs with probability $1$ for a stochastic fading channel. Then using
(\ref{eqgamma}), we have the following expression for the $(i,j)$th element of
$\gamma_m$
\beqa
\gamma_m^{i,j} & = & \frac{1}{N} \sum_{l=1}^P \sum_{n=0}^{N-1}
\frac{\alpha_l^{i,j}}{e^{-j2 \pi n/N} - \beta_l} e^{j2\pi mn/N} \nonumber \\
& = & \sum_{l=1}^{P} \alpha_l^{i,j} \sum_{n=0}^{N-1} \frac{e^{j2 \pi m n/N}}{N} 
\sum_{q=0}^\infty \left[e^{j2 \pi (q+1)n/N} \beta_l^q
{\mathbf{1}}_{|\beta_l| < 1} - \frac{e^{-j2 \pi qn/N}}{(\beta_l)^{q+1}}
{\mathbf{1}}_{|\beta_l| > 1} \right] \\
& = & \sum_{l=1}^{P} \alpha_l^{i,j} \sum_{q=0}^\infty
\frac{1}{N} \sum_{n=0}^{N-1} \left[ e^{j2 \pi (m+q+1)n/N} \beta_l^q
{\mathbf{1}}_{|\beta_l| < 1} - \frac{e^{j2 \pi (m-q)n/N}}{(\beta_l)^{q+1}}
{\mathbf{1}}_{|\beta_l| > 1} \right] \\
\label{eqdelta}
& = & \sum_{l=1}^{P} \alpha_l^{i,j} \sum_{q=0}^\infty
\left[ \delta_{q,(rN+N-1-m)} \beta_l^q {\mathbf{1}}_{|\beta_l| < 1} -
\delta_{q,(m+rN)} {\mathbf{1}}_{|\beta_l| > 1} \right] \\
& = & \sum_{l=1}^{P} \alpha_l^{i,j} \left[ \beta_l^{N-1-m} \sum_{r=0}^\infty \beta_l^{rN} 
{\mathbf{1}}_{|\beta_l| < 1} - \frac{1}{(\beta_l)^{m+1}} \sum_{r=0}^\infty
(\beta_l)^{-rN} {\mathbf{1}}_{|\beta_l| > 1} \right] \\
& = & \sum_{l=1}^{P} \alpha_l^{i,j} \Bigg[ \frac{\beta_l^{N-1-m}}{1 - \beta_l^{N}} 
{\mathbf{1}}_{|\beta_l| < 1} -
\frac{1}{(\beta_l)^{m+1}} \frac{1}{\left[1 - (\beta_l)^{-N}\right]}
{\mathbf{1}}_{|\beta_l| > 1} \Bigg] \\
& = & \sum_{l=1}^{P} \alpha_l^{i,j} \frac{\beta_l^{N-1-m}}{1 - \beta_l^{N}},
\enqa
where in (\ref{eqdelta}), $\delta_{i,j} = 1$ if $i=j$ and is zero otherwise,
${\mathbf{1}}_{\mbox{\scriptsize condition}}$ is the indicator function that is $1$ when the `condition' is true
and is zero otherwise. It follows from the above expression that the contribution of the pole $\beta_l$ to the
`Edge Effect' decreases exponentially as $\beta_l^{-m-1}$ as one moves inside the frame from its head for $|\beta_l|>1$,
and as $\beta_l^{N-m-1}$ as one moves inside the frame from its tail for $|\beta_l|<1$. For the case of MMSE SC-FDE where the
roots occur in conjugate reciprocal pairs, the `Edge Effect' is determined primarily by the
pair of roots closest to the unit circle that make it decay with the same rate from both the head and
the tail of the frame.

We also note that for $K > 0$,
\beq
R(e^{jw}) = e^{jw(L-1)} \det \left( H(e^{jw}) H^\dagger(e^{jw})  + KI \right) > 0,
~~~ \forall ~~~ w.
\enq
Hence for the MMSE SC-FDE, there is no pole on the unit circle and the edge effect will decay
exponentially as one moves within the frame.

\subsection{CP-less OFDM}

For the OFDM transmission, one can recover the symbol tones by taking the DFT
of the equalized signal recovered without the CP as in the previous section. The
SINR expression unlike the SC-FDE, doesn't have a closed form expression.

\section{Case of a two-path fading channel}

Let us consider the case of a two tap fading channel for $N_r = N_t = 1$
whose $z$-transform is given by
\beq
H(z) = h_0 + h_1 z^d,
\enq
where $d$ is an integer that denotes the delay of
the second path, and $h_0$ and $h_1$ are ${\cal CN}(0,1)$.
For the case of ZF SC-FDE, we look at the roots of $H(z)$ that have a
magnitude of $|h_0/h_1|^{1/d}$. The probability that there are roots that have the magnitude in the
interval $(1-\epsilon,1+\epsilon)$, with $\epsilon \in (0,1)$, is given by
\beqa
p_{\epsilon}
 & = & \mbox{E} \left\{ \mbox{Prob} \left[ |h_0|^{2/d} \in (|h_1|^{2/d}(1-\epsilon)^2, |h_1|^{2/d}(1+\epsilon)^2 | h_1 \right] \right\}\\
& = & \frac{(1+\epsilon)^{2d} - (1-\epsilon)^{2d}}{1 + (1-\epsilon)^{2d} + (1+\epsilon)^{2d} + (1-\epsilon^2)^{2d}}.
\enqa
For $d=1$, this simplifies to
\beq
p_{\epsilon} = \frac{\epsilon}{1 + \epsilon^4/4}.
\enq
For small $\epsilon$, $p_{\epsilon} \approx d \epsilon$. $p_{\epsilon}$ increases with $d$ since it makes the roots of $H(z)$
to be closer to the unit circle. This implies that the response of the inverse filter (see (\ref{eqgamma})) will take longer to die down.

For the case of the MMSE SC-FDE, we need to look at the roots of
\beqa
R(z) & = & z^d H(z) H^*\left(\frac{1}{z^*}\right) + K z^d \nonumber \\
& = & h_0 h_1^* + (|h_0|^2 + |h_1|^2 + K)z^d + h_0^* h_1 z^{2d}. \nonumber
\enqa
The roots of $R(z)$ are the $d$th roots of
\beq
\label{eqrho}
\rho_{1,2} = \frac{-\psi \pm \sqrt{\psi^2 - 4|h_0|^2 |h_1|^2}}{2 h_0^* h_1},
\enq
where $\psi = |h_0|^2 + |h_1|^2 + K$.
It is easily shown that $\rho_1 = 1/\rho_2^*$, $|\rho_2| > |\rho_1|$, and hence $|\rho_2| \geq 1$, and
\beqa
|\rho_2| & \geq & \max \left( \left| \frac{h_0}{h_1} \right|, \left| \frac{h_1}{h_0} \right| \right), \\
|\rho_1| & \leq & \min \left( \left| \frac{h_0}{h_1} \right|, \left| \frac{h_1}{h_0} \right| \right).
\enqa
This implies that the roots of $R(z)$ are farther away from the unit circle as compared to the roots of $H(z)$,
and hence the response of the inverse filter in (\ref{eqgamma}) dies faster for the
MMSE SC-FDE than the ZF SC-FDE.

One can simplify (\ref{eqrho}) to get
\beq
\label{eqh0}
|h_0|^2 - \left( |\rho_2| + \frac{1}{|\rho_2|} \right) |h_0||h_1| + (|h_1|^2 + K) = 0,
\enq
which admits a solution for $|h_0|$ only if
\beq
|h_1|^2 \geq \frac{4 |\rho_2|^2 K}{(1-|\rho_2|^2)^2}.
\enq
For $|\rho_2| = 1 + \epsilon$, where $\epsilon > 0$, this amounts to
\beq
|h_1|^2 \geq \left(\frac{1+\epsilon}{1+\epsilon/2}\right)^2 \frac{K}{\epsilon^2} > \frac{K}{\epsilon^2},
\enq
an event which occurs with the probability of less than $e^{-K/\epsilon^2}$. For $\epsilon < 0.5\sqrt{K}$ and
$\epsilon < 0.25\sqrt{K}$, the probability is less than $2\%$ and $0.000012\%$ respectively.
Hence there are no roots of $R(z)$ in $((1 + 0.25\sqrt{K})^{-1/d},(1+0.25\sqrt{K})^{1/d})$ with probability approaching
unity. Hence one can be sure with probability approaching unity
that the response of the inverse filter in (\ref{eqgamma}) decays
exponentially with the exponent of at least $(1+0.25\sqrt{K})^{1/d}$ from
the head and the tail. Hence one can choose $D$ large enough such that
$\xi_m$ in (\ref{xieq}) or more conveniently the upper bound to $|\xi_m|$ in (\ref{ubxi}) is
small.

If the second path is weaker than the first path i.e. has an average power of $\sigma^2$ where $\sigma^2 < 1$, then
there are no roots of $R(z)$ in $((1 + 0.25\sqrt{K}/\sigma)^{-1/d},(1+0.25\sqrt{K}/\sigma)^{1/d})$ with probability
less than $0.000012\%$. This makes the response of the inverse filter decrease faster than the case when both
paths have same average power. Note that if the second path has larger average power than the first path,
then we can view (\ref{eqh0}) as a quadratic equation in $|h_1|^2$ and follow the same arguments to get to
the same conclusion as above.

\begin{figure}[t]
\centering{
\includegraphics[height=4.3in,angle=90]{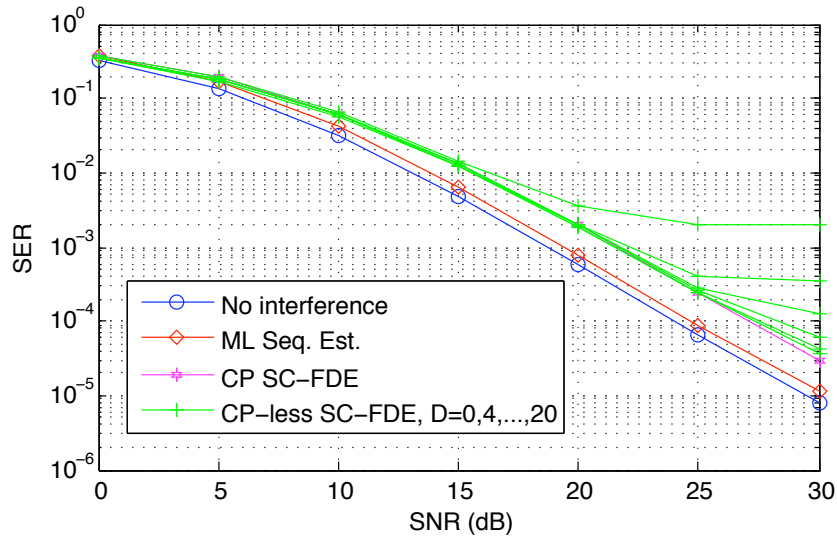}
}
\caption{Symbol Error Rate (SER) versus SNR curves for QPSK for the two-path fading channel with $d=1$, $D = 0,4,8,12,16,20$, and DFT frame size of $512$.}
\label{fig1}
\end{figure}

We plot the numerical results for this channel for QPSK modulation in Fig. \ref{fig1} with maximum 
likelihood sequence estimation using the Viterbi algorithm and the SC-FDE with or without CP. We
also plot  the case when there is no self-interference i.e. a flat fading
channel with the same average received power of unity as the ISI channel.
The size of the DFT frame is $512$. For the case of SC-FDE
with no CP, we consider the inner portion (quantized by parameter $D$) as
equalized in one frame and the exterior portion of the frame from both ends forms the interior
portion of some other frame. As one can see in Fig. \ref{fig1},
as $D$ increases, the performance of CP-less SC-FDE approaches that of CP SC-FDE. This matches
well with the earlier analytical observation that the SINR for the inner portion of the frame
without CP approaches that of the SINR with CP for SC-FDE.

\section{Conclusions}

It is shown that for SC-FDE without the cyclic prefix 
by increasing the length of the DFT and by discarding the symbols
on either ends of the DFT frame one can asymptotically obtain the same value of SINR as the
case when the cyclic prefix is present. We showed for a two-path fading channel
(whose inverse channel has a long tail) that the proposed method presents a method
to do away with the redundancy due to the cyclic prefix by a moderate increase
in the receiver complexity. The loss of redundancy could be used to achieve
high data rates due to increase in the number of channel uses.

\appendix
\section{Proof}
We can write using (\ref{ofdmrx})
\beq
\tilde{\mycal{R}} = \sqrt{N} \tilde{\mycal{H}} \mycal{F}^\dagger \mycal{Y} +
\tilde{\mycal{V}},
\enq
where $\tilde{\mycal{R}}$ $= [\tilde{r}_0 \cdots \tilde{r}_{N-1}]^T$,
$\mycal{F}$ is a $NN_t \times NN_t$ block diagonal unitary matrix given by
$\mycal{F} = \mbox{diag}\left[F,\cdots,F\right]$,
where $F$ is the $N \times N$ DFT unitary matrix,
$\tilde{\mycal{Y}}$ $= [\tilde{y}_0 \cdots \tilde{y}_{N-1}]^T$,
and $\tilde{\mycal{V}}$ $= [\tilde{\nu}_0 \cdots \tilde{\nu}_{N-1}]^T$.
Since we are interested in obtaining $Y$, we can write the ZF or MMSE pre-multiplying
matrix as (with $K = 0$ for ZF and $K = N_0$ for MMSE)
\beqan
\mycal{G} & = & \sqrt{N}
\left(  N \mycal{F} \tilde{\mycal{H}}^\dagger \tilde{\mycal{H}}
\mycal{F}^\dagger + K I \right)^{-1} \mycal{F} \tilde{\mycal{H}}^\dagger \\
& = & \sqrt{N} \mycal{F}
\left(  N  \tilde{\mycal{H}}^\dagger \tilde{\mycal{H}}
+ K I \right)^{-1} \tilde{\mycal{H}}^\dagger,
\enqan
which is a concatenation of ZF or MMSE in the frequency domain followed by DFT. Note
that $\mycal{G}$ $=\mbox{diag} [g_0, \cdots, g_{N-1}]$
is a block diagonal matrix, where
\beq
\tilde{g}_n =
\sqrt{N}
(N\tilde{h}_n^\dagger \tilde{h}_n + KI)^{-1} \tilde{h}_n^\dagger.
\enq

\end{document}